# *In-situ* synchrotron-based high energy X-ray diffraction study of the deformation mechanism of δ-hydride in a commercially pure titanium

Qing Tan [a,b,c, *], Zhiran Yan [a], Runguang Li [d], Yang Ren [e], Yandong Wang [a],

Baptiste Gault [b,c] Stoichko Antonov [b]

[a] State Key Laboratory for Advanced Metals and Materials, University of Science and Technology Beijing, Beijing 100083, China

[b] Department of Microstructure Physics and Alloy Design, Max-Planck-Institut für Eisenforschung GmbH, Max-Planck-Straße 1, 40237, Düsseldorf, Germany

[c] Department of Materials, Royal School of Mines, Imperial College, Prince Consort Road, London SW7 2BP, United Kingdom

[d] Department of Mechanical Engineering, Technical University of Denmark, DK-2800, Kgs. Lyngby, Denmark

[e] X-ray Science Division, Advanced Photon Source, Argonne National Laboratory, Argonne, IL 60439, USA

* corresponding author: q.tan@mpie.de

**Abstract:**

We investigated the deformation behavior of Grade 2 commercially pure titanium that was hydrogen charged to form hydrides by *in-situ* high energy X-ray diffraction. The results showed that high internal and interphase stresses generated within and around hydrides due to the volume expansion induced by the phase transformation could lead to the peak broadening. Hydride behaves as typical high strength but brittle secondary phase, which undertakes more elastic strain than matrix and generates cracks first. This study on the deformation mechanism of hydrides in titanium provides insight into the hydride behavior and hydrogen embrittlement in both titanium and zirconium.

Due to the excellent strength-to-weight ratio and corrosion resistance, titanium and its alloys are important structural materials for applications in industries ranging from aerospace to biomedical [1,2]. Despite exhibiting lower strength levels compared to other Ti alloys, the excellent corrosion resistance of commercially pure (CP), hexagonal close-packed (hcp) α-titanium makes it an attractive material for components subjected to a variety of aggressive environments, e.g., sea water, human body tissue and fluids and other aqueous solutions [3,4]. Titanium, however, has a high affinity for H [5,6], making it highly susceptible to hydrogen embrittlement, which can cause failure of CP titanium especially in high pH environments, or under high temperatures and H-pressures [7–10]. Depending on the environmental and loading conditions, different mechanisms for hydrogen embrittlement have been proposed, including the formation of brittle hydrides or hydrogen enhanced local plasticity [11–14]. One key factor for the embrittlement seems to be the low solubility of hydrogen in the α phase [15], which leads to the formation of brittle hydrides [11,16,17].

Many experimental and theoretical studies on the structure and properties of hydrides (in titanium and zirconium) have been carried thus far [18–24]. As explored by first-principle calculations [19,24], three types of hydrides can form in α-titanium depending on the hydrogen concentration: (i) the face-center tetragonal γ-TiH forms at lower H contents, which is widely considered as a metastable variant; (ii) the more stable face-centered cubic (fcc) δ-TiH$_x$, where x ranges from 1.5 to 2, forms at intermediate to high hydrogen contents; and (iii) ε-TiH$_2$, which is a tetragonally distorted form of δ. These hydrides typically form via H atoms randomly occupying the tetrahedral interstitial sites of the Ti lattice [24,25]. Numerous studies have reported on how these hydrides affect the deformation properties of titanium alloys, including strength and ductility [8,26], and fatigue behavior [20,21,27]. On the one hand, it is generally accepted that hydrides reduce ductility and fatigue life significantly. On the other hand, the hardness and the yield stress at room temperature are very sensitive to the hydrogen content and hardness measurements are often scattered [21]. For instance, most studies show that titanium or zirconium hydrides (γ, δ or ε) have a higher Young's modulus, hardness and yield strength than those of the parent matrix, but these decrease with increasing hydrogen content [6,11,28].

Most studies dealing with the interactions between the dislocations or twins of α titanium and the hydrides are *ex-situ* and utilize transmission electron microscopy (TEM) observations [8,27]. One of the challenges in studying hydrides behavior in CP titanium and Ti-based alloys is the formation and dissolution of new hydrides during the preparation of specimens by electro-polishing or room temperature focused ion beam (FIB) milling [29,30]. Whether these specific hydrides exhibit the same properties as those formed in bulk samples, and whether post-deformation sample preparation alters the hydrides-matrix deformation structure and interaction, remain open questions.

*In-situ* synchrotron-based high-energy X-ray diffraction (HE-XRD) can be a powerful method to investigate the deformation mechanisms and interactions between precipitates and the parent matrix. Hydride deformation in zirconium was recently investigated by synchrotron-based HE-XRD and the results indicated that a deformation-induced phase transformation from δ to ζ occurred during the early stage of tensile plastic deformation [31]. Although Zr and Ti exhibit similarities when it comes to hydrogen embrittlement, Zr has a much higher H solubility and its hydrides tend to grow to larger sizes [24,32]. In order to improve the understanding of hydrogen embrittlement in titanium alloys, and maybe also to zirconium alloys, we study the deformation of CP-Ti Grade 2 *in-situ* by HE-XRD, and report on the deformation behavior of the hydrides and its interaction with the parent Ti matrix.

A Grade 2 CP-Ti extruded bar with a diameter of 20 mm and a starting grain size of ~10 μm was used for this study. Dog-bone-shaped tensile specimens were extracted from the center of the bar parallel to the extrusion direction by electro-discharge machining, and were ground and polished to remove the surface oxide layer and ensure H absorption[33]. Hydrogen was introduced by cathodic charging in a 1 mol/L H$_2$SO$_4$ water solution, with a current density of 300 mA/cm$^2$, at room temperature for 3 days. Subsequently, both the hydrogen charged (Ti-H) and hydrogen-free (Ti) samples were annealed in vacuum at 623 K for 6 h. The orientation relationship between hydrides and the matrix was observed using electron backscatter diffraction (EBSD) on a ZEISS Supra 55 scanning electron microscope (SEM) from the grip sections of the deformed samples, and all imaging was performed in back scattered electron (BSE) mode. Room temperature *in-situ* HE-XRD tensile tests were performed at beamline 11-ID-C at the Advanced Photon Source, Argonne National Laboratory (APS, ANL). The experimental setup has been detailed in Ref. [31]. A monochromatic X-ray beam (0.5×0.5 mm$^2$) with an energy of 106 keV (wavelength 0.11725 Å) was used. The loading direction (LD) and transverse direction (TD) correspond to azimuth angles of 90° and 0°, respectively. The pseudo-Voigt function was employed to fit the diffraction peaks to calculate the peak position, integrated intensity and full width at half maximum (FWHM). The lattice strain ($\varepsilon_{hkl}$) of the titanium matrix and hydrides was obtained by calculating the lattice spacing change during deformation using the formula $\varepsilon_{hkl} = \frac{d_{hkl} - d_0}{d_0}$, where $d_{hkl}$ and $d_0$ are the interplanar spacing (d-spacing) under a specified applied stress and the at stress-free state, respectively. The modulus of the (220) hydride planes $E_{220}$ can be calculated approximately by $E_{(220)} = \frac{S_{(220)}}{S_{(a_1a_2a_3c)}} \times E_{(a_1a_2a_3c)}$, where $E_{(a_1a_2a_3c)}$ is the modulus of the $(a_1a_2a_3c)$ plane of titanium and $S_{(220)}$, $S_{(a_1a_2a_3c)}$ are the slope of elastic stage of lattice plane (220) and $(a_1a_2a_3c)$, respectively.

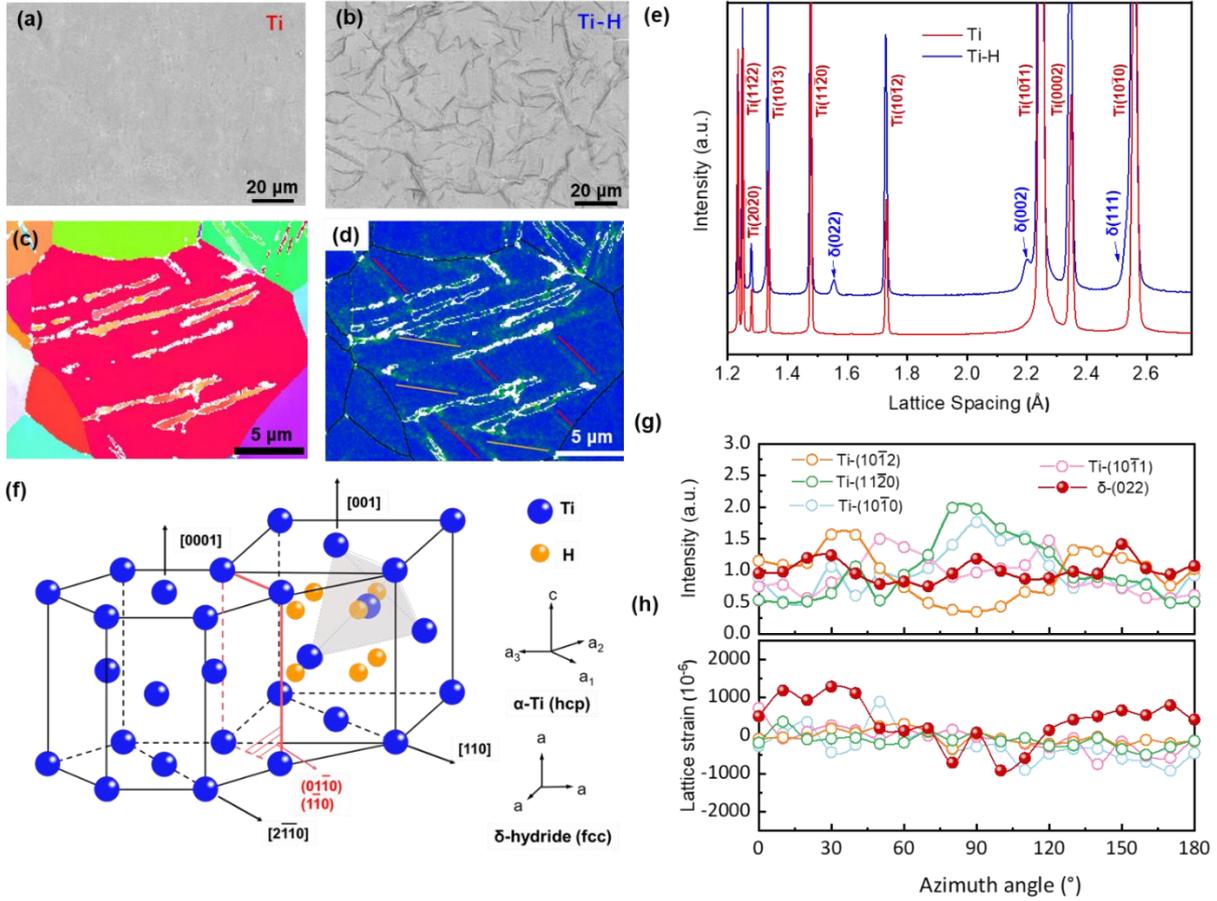

**Figure 1.** BSE images of the (a) Ti and (b) Ti-H samples; (c) orientation and (d) local misorientation of a representative grain in the Ti-H sample; (e) x-ray diffraction pattern of the pre-deformation Ti and Ti-H samples; (f) a schematic of the hydride and titanium matrix crystal structures and interface; (g) integrated intensity profiles from the x-ray diffraction Debye-Scherrer rings; and (h) calculated lattice strain along the azimuth angle (0~180°) on different lattice planes.

Representative images from the electropolished hydrogen-free (Ti) and hydrogen charged (Ti-H) CP-Ti samples are shown in Fig. 1a and Fig. 1b, respectively. Hydride plates with an average thickness of <1 μm and a length of ~10 μm can be seen only in the Ti-H sample. Both samples exhibit the same texture and average grain size of 20 μm, indicating that hydrogen charging does not change the grain size and texture of the matrix. Fig. 1c and d show the orientation and local misorientation maps within a grain of the Ti-H sample, respectively, where the hydrides are readily visible. The interface between these particles and the matrix is not well indexed because the lattice mismatch is accommodated by a high density of interfacial dislocations [35]. The epitaxial orientation relationship (OR) was determined as $(0001)_\alpha//(001)_\delta$ and $[1\bar{2}10]_\alpha//[110]_\delta$, which is consistent with previous reports [35]. Additionally, many bands of increased local misorientation with the same orientation as the hydrides, marked with red and yellow lines in Fig. 1d, can be observed in the matrix. These bands are related to the high internal stresses in the matrix arising from the volume expansion associated with hydride formation, and likely corresponds to sub-surface or hydrides removed during the specimen preparation.

Fig. 1e shows the X-ray diffraction patterns of the Ti and Ti-H samples. Only α phase peaks are present for the former, while the latter exhibits additional peaks indexed as δ-hydride with a calculated lattice parameter of 4.41 Å, consistent with the existing literature [35]. No peaks corresponding to γ-hydrides or other phases were indexed. As previously mentioned, the formation of the δ-hydrides leads to a volume expansion, and in this study, the unit cell volume of δ-hydrides was determined to be 84.9 Å$^3$, which is 20% larger than that of the matrix, 70.8 Å$^3$. Based on this information, a schematic of a hydride/titanium matrix interface structure is shown in Fig. 1f. The hydrides grow with a plate-like morphology where the $(2\bar{2}0)_\delta$ planes are coherent to the $(01\bar{1}0)_\alpha$ planes. The misfit in the habit plane parallel to the platelet, *i.e.* $(21\bar{1}0)_\alpha$ and $(2\bar{2}0)_\delta$, was calculated as 5.0% ($d_{(21\bar{1}0)_\alpha}$ = 1.4761 Å, $d_{(2\bar{2}0)_\delta}$ = 1.5541 Å ), while that for the $(0001)_\alpha$ and $(001)_\delta$, is -6.6% ($d_{(0001)_\alpha}$= 4.69Å, $d_{(001)_\delta}$ = 4.40 Å). The stress inside the hydrides parallel to the habit plane is mainly affected by the coherent misfit strain, which is not a simple compressive or tensile stress due to the superposition of positive and negative misfits of different planes. The misfit vertical to the platelet, *i.e.*, $(01\bar{1}0)_\alpha$ and $(1\bar{1}0)_\delta$, is more than 17% ($d_{(01\bar{1}0)_\alpha}$= 2.56Å, $d_{(1\bar{1}0)_\delta}$ = 3.02 Å), and the strain along the vertical habit plane direction is compressive, which is consistent with the TEM results [35]. Due to the overlap of the $(002)_\delta$ and $(111)_\delta$ with the $(10\bar{1}1)_\alpha$ and $(10\bar{1}0)_\alpha$ matrix peaks, only the $(220)_\delta$ peak was used for further analysis of the hydrides' deformation behavior.

The integrated intensity from X-ray diffraction Debye-Scherer rings for different lattice planes along the azimuth angle (0~180°) are presented in Fig. 1g. The titanium matrix has a strong preferred orientation of $<10\bar{1}0>$ // LD due to the extrusion. The lattice strain along the azimuth angle (0~180°) is shown in Fig. 1h, and the d-spacing of $(2\bar{2}0)_\delta$ along LD is smaller than along TD, since the hydride platelets have a preferred orientation normal to the LD and compressive stresses along the direction normal to the platelet, as discussed above, leading to a slightly larger average compressive stress along LD than TD in the hydrides. As the hydrides' size is much smaller than that of the matrix, the interphase stress affects the hydrides significantly more compared to just a part of the surrounding matrix, and so the lattice strain fluctuation for the hydrides is larger than all the planes of the titanium matrix, with a maximum reaching 2000 με.

The engineering stress-strain curves of the Ti and Ti-H specimens from the *in-situ* HE-XRD testing are shown in Fig. 2a. The yield stress of the Ti-H sample (558 MPa) is slightly lower than that of the Ti sample (572 MPa), and the hydrides significantly reduced the ductility, as cracks were formed from the hydrides - shown in Fig. 2b. The fracture morphologies of the two samples are shown in Fig. 2c and Fig. 2d. The Ti sample exhibits a typical ductile fracture with large, deep dimples, while the Ti-H sample has brittle cleavage fracture characteristics. The samples were also observed along the longitudinal sections near the fracture surfaces, as shown in Fig. 2e and Fig. 2f, and many small cracks can be seen in the Ti-H sample, while neither cracks nor voids were observed in the Ti sample. These small cracks, which are normal to the LD, are due to the brittle hydrides fracturing prior to the tearing of the metal matrix, consistent with observations in α+β Ti-6Al-4V[36].

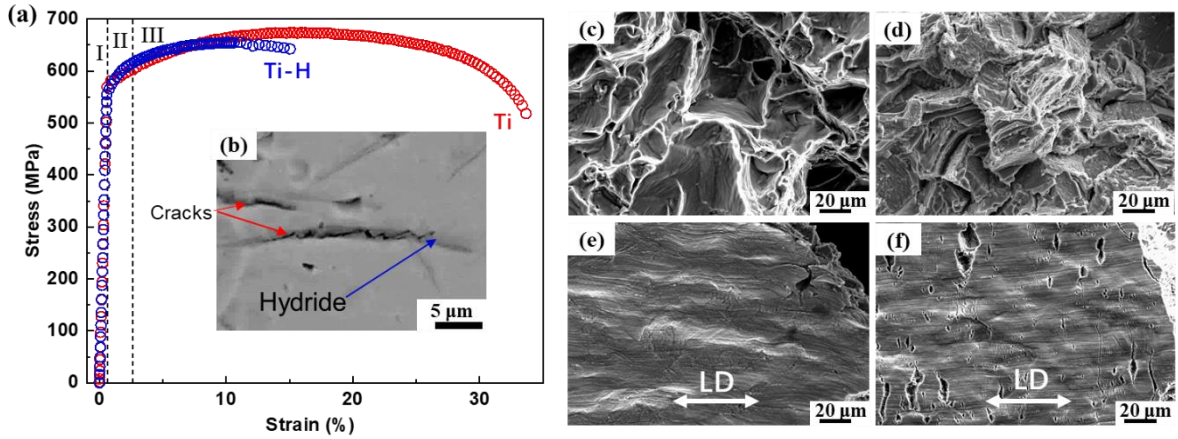

**Figure 2.** (a) Engineering tensile stress-strain curves of Ti and Ti-H. (b) The crack inside the hydrides after fracture. The morphology of fracture and surface near fracture of (c, d) Ti and (e, f) Ti-H.

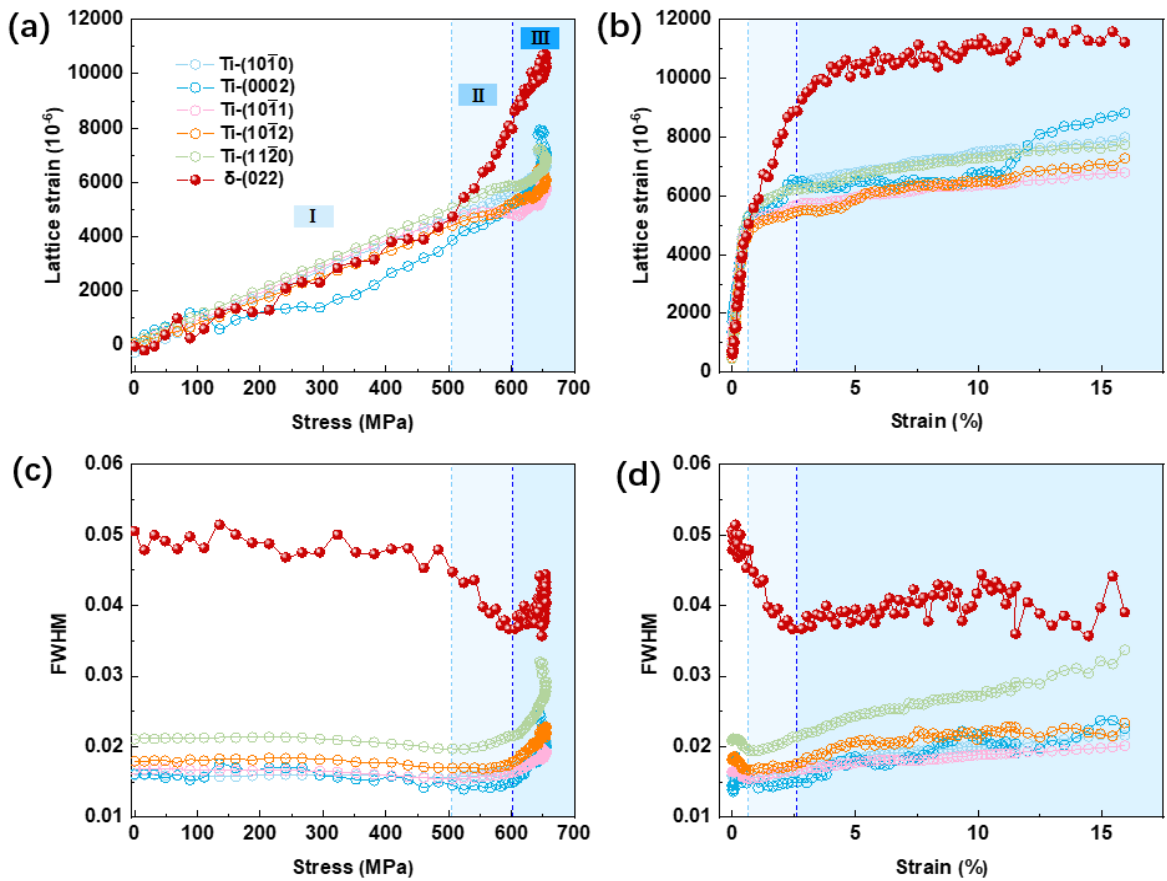

**Figure 3.** Evolution of lattice strain and FWHM along LD of α-titanium and δ-hydrides as a function of (a, c) engineering stress and (b, d) engineering strain.

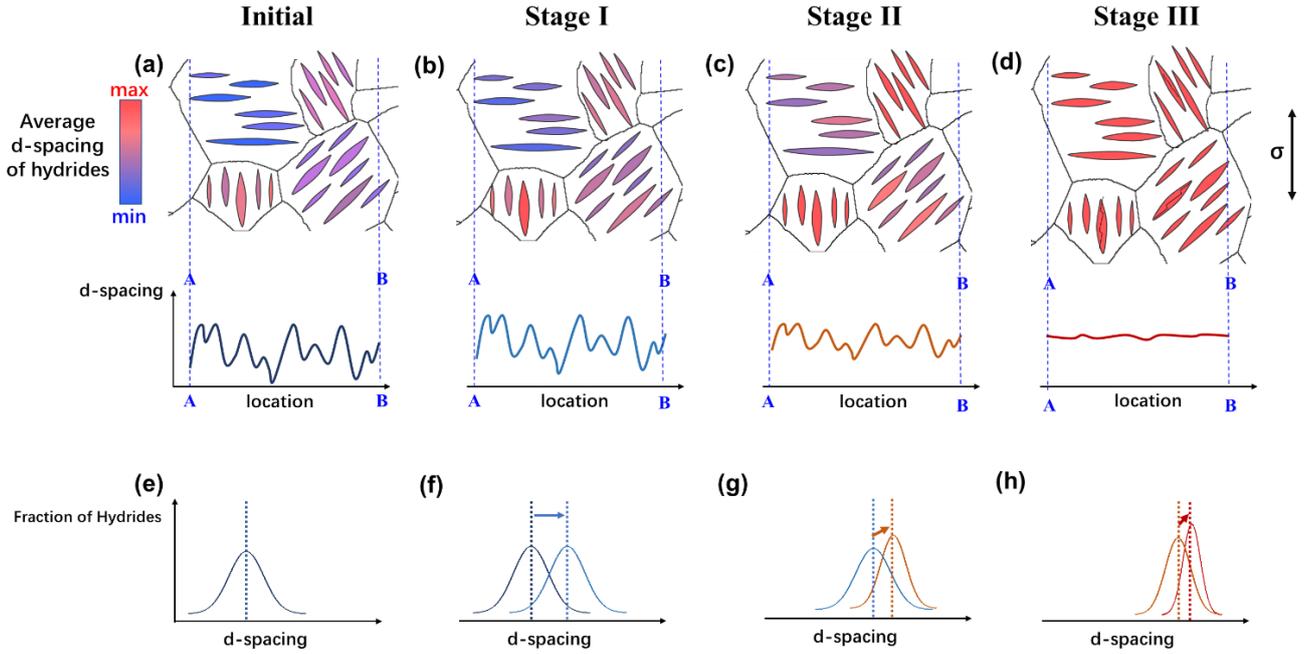

**Figure 4.** Schematics for the d-spacing of hydrides in different stages. (a) initial (b) Stage I (c) Stage II (d) Stage III. In the second row, the lines below are the schematics of d-spacing distribution in hydrides versus the location from A to B. In the third row, the schematics describe of d-spacing statistics distribution of all the hydrides in different stage (e) initial (f) Stage I (g) Stage II (h) Stage III, and the dot lines present the average d-spacing.

The lattice strain evolution along the LD of $(10\bar{1}0)_\alpha$, $(0002)_\alpha$, $(10\bar{1}1)_\alpha$, $(10\bar{1}2)_\alpha$, $(11\bar{2}0)_\alpha$ and $(022)_\delta$ of the Ti-H sample versus engineering stress and strain are plotted in Fig. 3a and Fig. 3b, respectively. All the curves of the lattice strain exhibit a linear relationship with the applied stress until the yielding point at approx. 500 MPa. During the entire tensile deformation until fracture, the δ-hydride peaks remain apparent and no additional peaks appear, indicating that the δ-hydrides are stable and there is no phase transformation. The evolution of the FWHM (full width at half maximum) along the LD versus engineering stress and strain are plotted in Fig. 3c and Fig. 3d, respectively. The FWHM of $(022)_\delta$ remains constant prior to 500 MPa and then decreases until 600 MPa. As the hydrides are considered brittle, they have little or no plastic deformation before fracture hence the plastic deformation behavior of the hydrides is not considered in this work. The whole tensile process can be divided into three stages according to the evolution of the lattice strain of the hydride and its FWHM, *viz.* (i) elastic deformation (Stage I), (ii) matrix yields and strain accommodation (Stage II), (iii) the matrix deforms plastically together with elastic deformation of the hydrides (Stage III). To illustrate the behavior during the different stages, a schematic for the d-spacing of the hydrides from the initial state and at the end of the different stages is shown in Fig. 4.

For the initial state, the value of FWHM of hydrides is much bigger than the matrix, which is similar to the x-ray diffraction results of hydrides in Zr[37]. On one hand, the hydrides have a preferred orientation based on that of the matrix and a compressive stress along the direction vertical to the hydride plate, as mentioned previously, leading to a significant variance in the measured d-spacing from LD in different grains, as shown in Fig. 4a. On the other hand, the lattice parameter could also vary with the local stoichiometry of δ-TiHx (1.5≤x < 2), also leading to a peak broadening of hydrides.

During stage I, the slope of the lattice strain curve for the hydride, as shown in Fig. 3a, is nearly the same as that of the matrix and the FWHM of all the lattice planes varies very little, Fig. 3c, indicating that the matrix and hydrides deform elastically together and the strain increases equally. The elastic modulus of titanium δ-hydrides has been determined theoretically [38] and experimentally [6,39–41], and reported to be in the range of 106 – 128 GPa, which is similar to that of α-titanium (115 GPa). The elastic modulus of $(220)_\delta$ from the synchrotron data was calculated to be 111 GPa, which is slightly higher than that of the matrix (96 GPa) and within the range previously reported. The average d-spacing increases due to the tensile loading and the d-spacing dispersion does not change in this stage, as shown in Fig. 4b and Fig. 4f.

During stage II, the matrix starts to yield under an applied stress of around 500 MPa, while the hydrides keep deforming elastically, manifesting as an increase of the $(022)_\delta$ planes lattice strain slope, Fig. 3a. The FWHM of the matrix starts to increase slightly due to the accumulation of dislocations during the plastic deformation, while that of the hydrides drops obviously, Fig. 3b. During this time, the matrix starts to tilt towards the tensile texture <10-10>//LD, same as the starting preferred orientation, so that the hydrides orientation distribution becomes tighter and consequently, the dispersion of LD d-spacings among grains decreases. Meanwhile, due to the local stresses imposed by the hydride, the surrounding titanium matrix can start to plastically deform at lower stress levels, *i.e.*, micro-yielding, which is also a common phenomenon in hcp metals[42–45]. The stresses imposed on the hydrides become more homogenous after the matrix begins to yield. These processes lead to an increase of the average d-spacing of the hydrides (continued loading) and a decrease of its dispersion (matrix micro-yielding) during stage II, Fig. 4c and Fig. 4g.

During stage III, all of the matrix is plastically deforming while the hydrides are still deforming elastically in conjunction with deformation strain of the matrix. In Fig. 3d, the slopes of all the curves during this stage are practically the same, which means that the deformation rate of the hydrides is approximately equal to the work hardening rate of the titanium matrix, *i.e.*, the hydrides do not deform plastically. The lattice strain of the hydrides reaches ~1.2 % before fracture. The fracture strength can be obtained by multiplying the lattice strain and lattice elastic modulus $E_{(220)_\delta}$, and is 1300 MPa. In brittle materials, the yielding strength should be approximately equal to the fracture strength. However, the reported yield stress of hydrides is only ~900 MPa [6]. The difference between the fracture strength and the yielding strength should be caused by the initial compressive stress of the hydrides, the average of which is ~400 MPa in this work. At the end of this stage, the stresses in the hydrides reach the yield point, brittle cracks nucleate from the hydrides and lead to the fracture of the whole sample, Fig. 4d and Fig. 4h.

In summary, the morphology, preferred orientation, and deformation behavior of face-centered-cubic hydrides in Grade 2 CP-Ti were investigated. The hydride is platelet-like with orientation relationship of $(0001)_\alpha//(001)_\delta$ and $[1\bar{2}10]_\alpha//[110]_\delta$, and the habit plane is $(2\bar{2}0)_\delta//(01\bar{1}0)_\alpha$, as confirmed by both EBSD and HE-XRD. High internal and interphase stresses were shown to be generated within and around hydrides due to the volume expansion induced by the phase transformation of α- Ti to δ-TiH$_{1.5}$. The deformation mechanism of δ-hydrides in titanium (considering the effect of the initial stresses) was investigated by *in-situ* synchrotron diffraction. In contrast to first principle calculations [46] and hydride deformation studies in zirconium [31], phase transformation of the hydrides was not observed during deformation. The δ-hydrides in the titanium bear larger strains than the matrix, but the strain distribution becomes more concentrated after the matrix yields. Peak broadening is typically considered to be mainly affected by the grain size and microstrains, as estimated by Williamson-Hall analysis [47]. However, in this case, the initially wide dispersion of the d-spacings of the hydrides has a great contribution to the peak broadening. This study provides insight into the hydride behavior and hydrogen embrittlement in both titanium and zirconium

**Acknowledgements:**

QT would like to acknowledge the National Natural Science Foundation of China (NSFC) (Grant No. 11805009). SA would like to acknowledge financial support from the Alexander von Humboldt Foundation. BG and QT are grateful for financial support from the EPSRC (under grant number EP/T01041X/1) and the ERC-CoG-SHINE-771602. The use of the Advanced Photon Source was supported by the US Department of Energy, Office of Science, Office of Basic Energy Sciences, under Contract No. DE-AC02- 06CH11357.

**Reference**

[1] G. Lütjering, J.C. Williams, Titanium, 2014.
[2] H. Warlimont, Titanium and Titanium Alloys, 2018.
[3] A.T. Sidambe, Materials 7 (2014) 8168–8188.
[4] I. v. Gorynin, Materials Science and Engineering A 263 (1999) 112–116.
[5] O.N. Senkov, J.J. Jonas, Metallurgical and Materials Transactions A: Physical Metallurgy and Materials Science 27 (1996) 1869–1876.
[6] J.J. Xu, H.Y. Cheung, S.Q. Shi, Journal of Alloys and Compounds 436 (2007) 82–85.
[7] K. Takashima, K. Yokoyama, K. Asaoka, J. Sakai, Journal of Alloys and Compounds 431 (2007) 203–207.
[8] C.Q. Chen, S.X. Li, H. Zheng, L.B. Wang, K. Lu, Acta Materialia 52 (2004) 3697–3706.
[9] J. Huez, X. Feaugas, A.L. Helbert, I. Guillot, M. Clavel, Metallurgical and Materials Transactions A: Physical Metallurgy and Materials Science 29 (1998) 1615–1628.
[10] V. Madina, I. Azkarate, International Journal of Hydrogen Energy 34 (2009) 5976–5980.
[11] C.L. Briant, Z.F. Wang, N. Chollocoop, Corrosion Science 44 (2002) 1875–1888.
[12] E. Tal-Gutelmacher, D. Eliezer, Journal of Alloys and Compounds 404–406 (2005) 621–625.
[13] A.M. Alvarez, I.M. Robertson, H.K. Birnbaum, Acta Materialia 52 (2004) 4161–4175.
[14] D.S. Shih, I.M. Robertson, H.K. Birnbaum, Acta Metallurgica 36 (1988) 111–124.
[15] A. San-Martin, F.D. Manchester, Bulletin of Alloy Phase Diagrams 8 (1987) 30–42.
[16] M.I. Luppo, A. Politi, G. Vigna, Acta Materialia 53 (2005) 4987–4996.
[17] J. Wen, N. Allain, E. Fleury, Materials Characterization 121 (2016) 139–148.
[18] Z. Li, P. Ou, N. Sun, Z. Li, A. Shan, Materials Letters 105 (2013) 16–19.
[19] P.A.T. Olsson, M. Mrovec, M. Kroon, Acta Materialia 118 (2016) 362–373.
[20] C.Q. Chen, S.X. Li, Materials Science and Engineering A 387–389 (2004) 470–475.
[21] C.Q. Chen, S.X. Li, K. Lu, Acta Materialia 51 (2003) 931–942.
[22] P.A.T. Olsson, J. Blomqvist, C. Bjerkén, A.R. Massih, Computational Materials Science 97 (2015) 263–275.
[23] H.J. Liu, L. Zhou, P. Liu, Q.W. Liu, International Journal of Hydrogen Energy 34 (2009) 9596–9602.
[24] E. Conforto, I. Guillot, X. Feaugas, Philosophical Transactions of the Royal Society A: Mathematical, Physical and Engineering Sciences 375 (2017).
[25] D. v. Schur, S.Y. Zaginaichenko, V.M. Adejev, V.B. Voitovich, A.A. Lyashenko, V.I. Trefilov, International Journal of Hydrogen Energy 21 (1996) 1121–1124.
[26] P.E. Irving, C.J. Beevers, Metallurgical Transactions 2 (1971) 613–615.
[27] C.Q. Chen, S.X. Li, K. Lu, Philosophical Magazine 84 (2004) 29–43.
[28] C.P. Liang, H.R. Gong, International Journal of Hydrogen Energy 35 (2010) 3812–3816.
[29] Y. Chang, W. Lu, J. Guénolé, L.T. Stephenson, A. Szczpaniak, P. Kontis, A.K. Ackerman, F.F. Dear, I. Mouton, X. Zhong, S. Zhang, D. Dye, C.H. Liebscher, D. Ponge, S. Korte-Kerzel, D. Raabe, B. Gault, Nature Communications 10 (2019) 1–10.
[30] R. Ding, I.P. Jones, Journal of Electron Microscopy 60 (2011) 1–9.

[31] S. Li, Y. Wang, Z. Che, G. Liu, Y. Ren, Y. Wang, Acta Materialia 140 (2017) 168–175.
[32] J.S. Bradbrook, G.W. Lorimer, N. Ridley, Journal of Nuclear Materials 42 (1972) 142–160.
[33] J. Kim, D. Hall, H. Yan, Y. Shi, S. Joseph, S. Fearn, R.J. Chater, D. Dye, C.C. Tasan, Acta Materialia 220 (2021).
[34] K.D. Liss, D. Qu, K. Yan, M. Reid, Advanced Engineering Materials 15 (2013) 347–351.
[35] H. Numakura, M. Koiwa, O. Metals, 32 (1984) 1799–1807.
[36] J. Kim, E. Plancher, C.C. Tasan, Acta Materialia 188 (2020) 686–696.
[37] M.A. Vicente Alvarez, J.R. Santisteban, P. Vizcaíno, G. Ribárik, T. Ungar, Acta Materialia 117 (2016) 1–12.
[38] B. Modulus, 1558 (1989) 1553–1558.
[39] Y. Taniyama, H. Cho, M. Takemoto, Changes 25 (2007) 157–165.
[40] M.I. Luppo, A. Politi, G. Vigna, Acta Materialia 53 (2005) 4987–4996.
[41] D. Setoyama, J. Matsunaga, H. Muta, M. Uno, S. Yamanaka, Journal of Alloys and Compounds 381 (2004) 215–220.
[42] W.D. Brentnall, W. Rostoker, Acta Metallurgica 13 (1965) 187–198.
[43] A.G. Zhou, S. Basu, M.W. Barsoum, Acta Materialia 56 (2008) 60–67.
[44] A. Abel, H. Muir, Acta Metallurgica 21 (1973) 99–105.
[45] S.J. Bates, D.J. Bacon, Microyielding in Alpha Titanium, 1980.
[46] Z. A. Matysina, D. v. Shchur, (2001).
[47] G.K. Williamson, W.H. Hall, Acta Metallurgica 1 (1953) 22–31.